# Filter Bank Fusion Frames

Amina Chebira, *Member, IEEE,* Matthew Fickus, *Member, IEEE,* and Dustin G. Mixon

*Abstract*—In this paper we characterize and construct novel oversampled filter banks implementing fusion frames. A fusion frame is a sequence of orthogonal projection operators whose sum can be inverted in a numerically stable way. When properly designed, fusion frames can provide redundant encodings of signals which are optimally robust against certain types of noise and erasures. However, up to this point, few implementable constructions of such frames were known; we show how to construct them using oversampled filter banks. In this work, we first provide polyphase domain characterizations of filter bank fusion frames. We then use these characterizations to construct filter bank fusion frame versions of discrete wavelet and Gabor transforms, emphasizing those specific finite impulse response filters whose frequency responses are well-behaved.

## I. INTRODUCTION

Frames play an increasing role in many signal processing applications [1], [2], [3]. In particular, they have found their niche in the engineering community through oversampled filter banks (FBs) that implement frame operators [4], [5]. The work presented here is most closely related to [6], which investigates the types of FB frames that are optimal for robust transmission, and introduces the concept of strongly uniform frames, proving their resilience against erasures. While many constructions of oversampled FBs exist, the design of nice oversampled FBs — noise and erasure robust FBs with simple reconstruction formulas, and built from few-tap filters with good frequency responses — remains a nontrivial problem. The aim of this paper is to benefit from recent developments in frame theory, namely to use fusion frames, to better solve this problem. Fusion frames were introduced in [7], where they were called frames of subspaces. They have proven useful in different settings such as packet encoding [8] and vector estimation from noisy measurements [9]. In this work, we combine fusion frames concepts with FBs, deriving characterizations and constructions of FB fusion frames.

Riding on the success of orthogonal FBs, such as those implementing discrete wavelet transforms, FB frames emerged as a necessary tool to efficiently implement redundant transforms. The fundamental works on FB frames [4], [5] characterize translation-invariant frames in $\ell^2(\mathbb{Z})$ in terms of the polyphase matrices of the corresponding FBs. In particular, FB frames are characterized in [5], while [4] derives the optimal frame bounds of a FB in terms of the singular values of the polyphase matrix. One of the most popular examples of FB frames is the cosine modulated FB introduced in [10]. For other examples, see [1], [2] and the references therein. The challenge in designing nice FB frames lies in the myriad of constraints they must satisfy: to combat noise, their frame bounds should be close to each other; to be efficiently implementable, the filters should have a low number of taps; in terms of frequency response, the filters should have good frequency selectivity as well as high stop band attenuation. Even when restated in the polyphase domain, the problem of designing such nice FBs remains a difficult subject of interest [11], [12], [13], [14].

Frames have become increasingly popular as their redundancy became a major asset in many applications. This redundancy is often used to buy stability, resilience to noise and robustness against channel erasures. For instance, the optimality of frames has been demonstrated in quantization set ups [3], as well as some source coding schemes. In particular, tight frames were proven to be robust to noise, and unit-norm tight frames (UNTFs) to be resilient to one or more erasures [15], [16], [17], [18]. The generalizations to resilience against packet erasures came later with the introduction of a new class of frames called fusion frames. Here, individual frame coefficients are generalized into projections onto subspaces. Fusion frames have already found applications to packet encoding, with their robustness against erasures being studied in [8], [9]. Indeed, in [9], it is shown that the class of fusion frames which, for random signals, are optimally robust against noise and erasures correspond to optimal Grassmannian packings, termed equidistance tight fusion frames. Fusion frames have also been used in other applications, such as modeling feature extraction [19], [20] and sparse recovery [21].

In this paper, we introduce and study a new class of FBs named *filter bank fusion frames (FBFF)*. Of particular interest is the class of tight FBFFs, which we characterize in terms of their polyphase representations and show that they correspond to generalized UNTFs. We also present ways of constructing FBFFs and tight FBFFs. One example of such a design involves taking products of paraunitary polyphase matrices with UNTFs. This is exactly the same design process that was used to construct strongly uniform tight frames in [6]. Indeed, strongly uniform tight frames are a special case of tight FBFFs; an alternative generalization of such frames, namely totally finite impulse response (FIR) FBs, is considered in [22]. Explicit constructions of strongly uniform tight frames are presented in [23], where the authors design these frames in a manner analogous to non-downsampled FBs. In fact, the notion of FBFFs can be viewed as a generalization of redundant discrete wavelet transforms with non-downsampled

A. Chebira is with the Audiovisual Communications Laboratory, School of Computer and Communication Sciences, Ecole Polytechnique Fédérale de Lausanne, 1015 Lausanne, Switzerland, email: amina.chebira@epfl.ch.

M. Fickus is with the Department of Mathematics and Statistics, Air Force Institute of Technology, Wright-Patterson Air Force Base, OH 45433, USA. This work supported by AFOSR F1ATA09125G003. The views expressed in this article are those of the authors and do not reflect the official policy or position of the United States Air Force, Department of Defense, or the U.S. Government.

D. G. Mixon is with the Program in Applied and Computational Mathematics, Princeton University, Princeton, NJ 08544, USA.

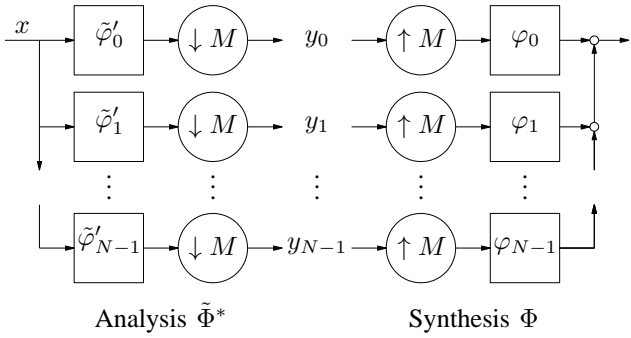

Fig. 1. An $N$-channel filter bank with a downsampling rate of $M$. A perfect reconstruction filter bank satisfies $\Phi\tilde{\Phi}^* = \mathrm{I}$. In particular, if the frame corresponding to $\Phi$ is tight, then there exists $A > 0$ such that $\Phi\Phi^* = A\mathrm{I}$, and so we may take $\tilde{\Phi} = \frac{1}{A}\Phi$.

FBs of integer redundancy, such as the ones presented in [24] and various other works; we pursue this vein, using traditional design methods, such wavelet and Gabor transforms, to build new examples of FBFFs. In particular, we build discrete wavelet and Gabor fusion frames, and in each case, design FIR filters that have good frequency responses.

In the next section, we introduce basic, previously understood concepts of FBs, frames and fusion frames. Then, in Section III, we characterize FBFFs in terms of their polyphase matrices, and provide several basic constructions of tight FBFFs. The remainder of the paper is devoted to constructing tight FBFFs whose filters have good frequency responses: discrete wavelet FBFFs are considered in Section IV, while Section V focuses on discrete Gabor FBFFs.

## II. BACKGROUND

In this section, we establish our notation, and discuss the previously-known results that we will need in order to characterize and construct FBFFs.

### A. Oversampled filter banks and frames

We consider oversampled FBs like those depicted in Fig. 1, in which each of the $N$ channels is downsampled by a factor of $M$, where $M < N$. So as to be more applicable to real-world problems, we restrict ourselves to finite-dimensional spaces of discrete periodic signals, namely:

$$\ell(\mathbb{Z}_P) := \{y : \mathbb{Z} \to \mathbb{C} \mid y[p+P] = y[p],\ p \in \mathbb{Z}\},$$

where $P$ is some positive integer. Equivalently, these are signals indexed by $\mathbb{Z}_P := \{0, \ldots, P-1\}$, the finite group of integers modulo $P$. In particular, given $N$ filters $\{\varphi_n\}_{n=0}^{N-1}$ in $\ell(\mathbb{Z}_{MP})$, the corresponding *synthesis FB* produces a single signal of period $MP$ from $N$ signals $y_n$ of period $P$: defined as an operator, we have $\Phi : [\ell(\mathbb{Z}_P)]^N \to \ell(\mathbb{Z}_{MP})$,

$$\Phi\{y_n\}_{n=0}^{N-1} = \sum_{n=0}^{N-1} \varphi_n * (\uparrow_M y_n), \quad (1)$$

where $\uparrow_M$ denotes upsampling by $M$, and each $y_n \in \ell(\mathbb{Z}_P)$ is the input of the $n$th channel of the synthesis FB. The adjoint of such an operator is $\Phi^* : \ell(\mathbb{Z}_{MP}) \to [\ell(\mathbb{Z}_P)]^N$,

$$\Phi^* x = \{\downarrow_M (\varphi'_n * x)\}_{n=0}^{N-1}. \quad (2)$$

where $\downarrow_M$ is downsampling by $M$, and $\varphi'_n[k] := (\varphi_n[-k])^*$ is the *involution* of $\varphi_n$. A *perfect reconstruction* FB consists of two sets of filters $\{\varphi_n\}_{n=0}^{N-1}$ and $\{\tilde{\varphi}_n\}_{n=0}^{N-1}$ whose corresponding synthesis and *analysis* FBs $\Phi$ and $\tilde{\Phi}^*$ satisfy $\Phi\tilde{\Phi}^* = \mathrm{I}$.

*Frame theory* provides a mechanism for studying the noise robustness of these transforms and, for a given synthesis FB $\Phi$, constructing a corresponding analysis FB $\tilde{\Phi}^*$. Specifically, letting $\mathbb{I}$ be a countable indexing set, a sequence of vectors $F = \{f_i\}_{i\in\mathbb{I}}$ lying in some Hilbert space $\mathbb{H}$ is a *frame* for $\mathbb{H}$ if there exists *frame bounds* $0 < A \leq B < \infty$ such that:

$$A\|f\|^2 \leq \sum_{i\in\mathbb{I}} |\langle f, f_i\rangle|^2 \leq B\|f\|^2,\ \forall f \in \mathbb{H}. \quad (3)$$

More generally, $\{f_i\}_{i\in\mathbb{I}}$ is a *Bessel sequence* if at least the upper bound in (3) is satisfied; in this case, its *synthesis operator* $F : \ell^2(\mathbb{I}) \to \mathbb{H}$ and corresponding adjoint $F^* : \mathbb{H} \to \ell^2(\mathbb{I})$ are both guaranteed to be linear and bounded:

$$Fg := \sum_{i\in\mathbb{I}} g(i)f_i, \quad (F^*f)(i) = \langle f, f_i\rangle. \quad (4)$$

The connection between frames and oversampled FBs is the fact that every synthesis FB (1) and its adjoint (2) are particular types of (frame) synthesis operators and their adjoints (4), respectively. Indeed, letting $\mathbb{H} = \ell(\mathbb{Z}_{MP})$, $\mathbb{I} = \mathbb{Z}_N \times \mathbb{Z}_P$, and identifying $\ell^2(\mathbb{I})$ with $[\ell(\mathbb{Z}_P)]^N$, one can show that the synthesis operator of $\Phi = \{\mathcal{T}^{Mp}\varphi_n\}_{n=0,\ p=0}^{N-1,\ P-1}$, namely:

$$\Phi\{y_n\}_{n=0}^{N-1} = \sum_{n=0}^{N-1}\sum_{p=0}^{P-1} y_n[p]\mathcal{T}^{Mp}\varphi_n,$$

is in fact identical to the synthesis FB (1), where $\mathcal{T}$ is the translation operator: $(\mathcal{T}^{k'}x)[k] := x[k-k']$. That is, the synthesis FB $\Phi$ is the synthesis operator of the $M$-translates of the filters $\{\varphi_n\}_{n=0}^{N-1}$. In this setting, the concept of a perfect reconstruction FB corresponds to finding a frame $F = \{f_i\}_{i\in\mathbb{I}}$ and a *dual frame* $\tilde{F} = \{\tilde{f}_i\}_{i\in\mathbb{I}}$ so that $F\tilde{F}^* = \mathrm{I}$, that is:

$$f = \sum_{i\in\mathbb{I}} \langle f, \tilde{f}_i\rangle f_i,\ \forall f \in \mathbb{H}.$$

However, not every frame, nor every perfect reconstruction FB, is created equal. Indeed, the whole point of frame theory is that certain transforms $F$ will be more stable and will perform better than others in the presence of noise, as indicated by their condition number. To clarify, the frame condition (3) is equivalent to having $A\mathrm{I} \leq FF^* \leq B\mathrm{I}$, where $FF^*$ is the *frame operator*:

$$FF^* = \sum_{i\in\mathbb{I}} f_i f_i^*. \quad (5)$$

When $\mathbb{H}$ is finite-dimensional, this implies that the least and greatest eigenvalues of $FF^*$ are the optimal expressions for the frame bounds $A$ and $B$, respectively, making $F$'s condition number $\sqrt{B/A}$. As such, for a given application, one typically tries to construct frames $F$ whose frame bounds $A$ and $B$ are as close as possible, subject to design constraints. Here, the best possible situation is when $F$ is a *tight frame*, that is, when $A = B$, implying $FF^* = A\mathrm{I}$. Tight frames are especially valuable in the FB setting: in general, any frame $F$ has a *canonical dual* $\tilde{F} = (FF^*)^{-1}F$ which satisfies $F\tilde{F}^* = \mathrm{I}$; however, in the FB setting, it is often the case that the canonical dual of FIR filters are no longer FIR; tight FBs are an exception to this rule, as $\tilde{\varphi} = \frac{1}{A}\varphi$.

*B. Fusion frames*

The purpose of this paper is to investigate FBs corresponding to a new class of frames, dubbed *fusion frames*, which are particularly suited to certain applications. Such frames arose as a generalization of *unit norm frames*, that is, frames $\{f_i\}_{i \in \mathbb{I}}$ for which $\|f_i\| = 1$ for all $i \in \mathbb{I}$. Of particular interest are UNTFs — unit norm frames that are also tight — which provide optimal robustness against noise and certain types of erasures [15], [16]. Intuitively, the value of UNTFs is that, like orthonormal bases, they provide isometric decompositions of signals in terms of components of equal weight. However, unlike orthonormal bases, UNTFs are nontrivial to construct: the easiest examples are the *harmonic tight frames* obtained by truncating a discrete Fourier transform (DFT) matrix; other examples are given in [25], [26]. From an operator-theoretic point of view, the significance of unit norm frames is that each of the summands $f_i f_i^*$ of the frame operator (5) is a rank-1 projection operator. Fusion frames generalize this idea so as to permit projections of higher rank.

To be precise, a collection of orthogonal projection operators $\{\Pi_k\}_{k \in \mathbb{K}}$ is a *fusion frame* if its *fusion frame operator* $\sum_{k \in \mathbb{K}} \Pi_k$ satisfies $A\mathrm{I} \leq \sum_{k \in \mathbb{K}} \Pi_k \leq B\mathrm{I}$ for some $0 < A \leq B < \infty$. Fig. 2 depicts an example of a tight fusion frame in $\mathbb{R}^3$ containing four rank-2 projections. These projections are onto the four orthogonal complements of the vertices of a tetrahedron. Here, $\sum_{k=0}^{3} \Pi_k = A\mathrm{I}$, where $A$ can be computed as the redundancy of the frame as $A = (4 \times 2)/3 = 8/3$. Though superficially a generalization of the concept of a unit norm frame, fusion frames are actually a special case of them: for each $k \in \mathbb{K}$, letting $F_k = \{f_{k,i}\}_{i \in \mathbb{I}_k}$ be an orthonormal basis for the range of $\Pi_k$, we classically know that $\Pi_k = F_k F_k^*$ and so:

$$\sum_{k \in \mathbb{K}} \Pi_k = \sum_{k \in \mathbb{K}} F_k F_k^* = \sum_{k \in \mathbb{K}} \sum_{i \in \mathbb{I}_k} f_{k,i} f_{k,i}^*. \quad (6)$$

That is, any fusion frame operator may be regarded as the (traditional) frame operator (5) of a doubly-indexed sequence $\{f_{k,i}\}_{k \in \mathbb{K}, i \in \mathbb{I}_k}$ where each subsequence $\{f_{k,i}\}_{i \in \mathbb{I}_k}$ is orthonormal. Put another way, a unit norm frame is a fusion frame if its elements may be partitioned into subsequences, each of which being an orthonormal set. From this perspective, any unit norm frame which has some mutually orthogonal elements, like many traditional wavelet and Gabor frames, is an example of a fusion frame. The purpose of this work is to better understand and exploit the mechanisms by which these orthogonality relationships occur.

Like traditional frames, the value of fusion frames is that they permit redundant, stable and noise-robust decompositions of signals. But whereas a frame $F = \{f_i\}_{i \in \mathbb{I}}$ breaks a signal $f$ into a possibly unorganized collection of frame coefficients $F^*f = \{\langle f, f_i \rangle\}_{i \in \mathbb{I}}$, a fusion frame provides a more hierarchical decomposition, breaking $f$ into packets $\{\Pi_k f\}_{k \in \mathbb{K}}$, each of which is itself a collection of coefficients, namely $\Pi_k = F_k F_k^*$, where $F_k^* f = \{\langle f, f_{k,i} \rangle\}_{i \in \mathbb{I}_k}$. For example, equal rank tight fusion frames, namely sequences of equal rank projections whose fusion frame operators (6) are scalar multiples of the identity, provide isometric redundant decompositions of signals in terms of packets of equal weight. Such frames are natural generalizations of UNTFs; though their existence has only recently been characterized [26], they are known to be optimal with respect to the erasure of entire packets [8], [9].

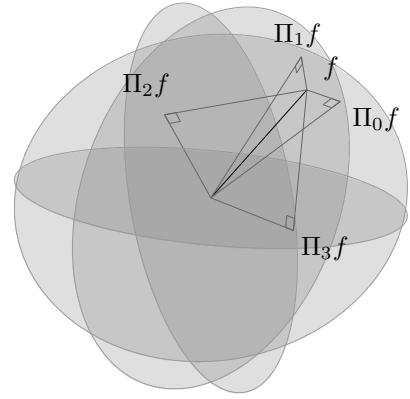

Fig. 2. A tight fusion frame consisting of four orthogonal projection operators $\{\Pi_k\}_{k=0}^{3}$ over $\mathbb{R}^3$, each of which has rank 2. The four corresponding column spaces are the two-dimensional orthogonal complements of the vertices of a tetrahedron; the unit disk in each of these spaces is depicted. Any given vector $f$ in $\mathbb{R}^3$ may be nonorthogonally decomposed in terms of its projections onto these four subspaces: $f = \frac{3}{8}(\Pi_0 f + \Pi_1 f + \Pi_2 f + \Pi_3 f)$.

III. CHARACTERIZING FILTER BANK FUSION FRAMES

This article focuses on when $\Phi$ corresponds to a *fusion* frame. In this section, we emphasize the special case when each of the $N$ channels of $\Phi\Phi^*$ is an orthogonal projection operator; such frames have been shown [8] to be optimally robust against the complete loss of data from any given single channel. In this section, we characterize FBFFs in terms of their polyphase matrices, and then use this characterization to provide several basic constructions of tight FBFFs.

*A. Polyphase Matrices of Filter Bank Fusion Frames*

Here, we combine the well-known polyphase characterizations of the orthogonality of translates [27] and the frame bounds of a FB [4] into a new characterization of FBFFs. Along the way, we recast these ideas from the infinite-dimensional $\ell^2(\mathbb{Z})$ setting into the more applicable, finite-dimensional $\ell(\mathbb{Z}_{MP})$ setting. In particular, we define the $M \times 1$ *polyphase vector* $\boldsymbol{\varphi}(z)$ of some filter $\varphi \in \ell(\mathbb{Z}_{MP})$ as:

$$\boldsymbol{\varphi}(z) := \begin{bmatrix} \varphi^{(0)}(z) \\ \vdots \\ \varphi^{(M-1)}(z) \end{bmatrix}, \quad \varphi^{(m)}(z) := \sum_{p=0}^{P-1} \varphi[m + Mp] z^{-p}, \quad (7)$$

where, for any $m = 0, \ldots M - 1$, $\varphi^{(m)}$ is termed the *mth polyphase component* of $\varphi$. Since $\varphi[m + Mp]$ is $P$-periodic in $p$, these entries are technically *cyclic polynomials*: they lie in the ring $\mathbb{C}[z]/\langle z^P - 1 \rangle$ of complex polynomials whose exponents are integers modulo $P$. The *polyphase space* $\mathbb{P}_{M,P}$ is the set of all possible such $\boldsymbol{\varphi}(z)$'s, and is a Hilbert space under the inner product:

$$\langle \boldsymbol{\varphi}(z), \boldsymbol{\psi}(z) \rangle_{\mathbb{P}_{M,P}} := \frac{1}{P} \sum_{p=0}^{P-1} \langle \boldsymbol{\varphi}(\mathrm{e}^{\frac{2\pi\mathrm{j}p}{P}}), \boldsymbol{\psi}(\mathrm{e}^{\frac{2\pi\mathrm{j}p}{P}}) \rangle_{\mathbb{C}^M}$$

$$= \frac{1}{P} \sum_{p=0}^{P-1} \sum_{m=0}^{M-1} \varphi^{(m)}(\mathrm{e}^{\frac{2\pi\mathrm{j}p}{P}}) (\psi^{(m)}(\mathrm{e}^{\frac{2\pi\mathrm{j}p}{P}}))^*.$$

In fact, using properties of the DFT, one can show that the mapping $\varphi \mapsto \boldsymbol{\varphi}(z)$, a discrete Zak transform [28], is unitary with $\langle \varphi, \psi \rangle_{\ell(\mathbb{Z}_{MP})} = \langle \boldsymbol{\varphi}(z), \boldsymbol{\psi}(z) \rangle_{\mathbb{P}_{M,P}}$ for any $\varphi, \psi \in \ell(\mathbb{Z}_{MP})$. Moreover, this fact, along with the easily verified relation $\mathcal{T}^{Mp} \varphi \mapsto z^{-p} \boldsymbol{\varphi}(z)$, yields that for any $x, \varphi \in \ell(\mathbb{Z}_{MP})$,

$$\langle x, \mathcal{T}^{Mp} \varphi \rangle_{\ell(\mathbb{Z}_{MP})} = \langle \mathbf{x}(z), z^{-p} \boldsymbol{\varphi}(z) \rangle_{\mathbb{P}_{M,P}}$$

$$= \frac{1}{P} \sum_{p'=0}^{P-1} \langle \mathbf{x}(\mathrm{e}^{\frac{2\pi\mathrm{j}p'}{P}}), \boldsymbol{\varphi}(\mathrm{e}^{\frac{2\pi\mathrm{j}p'}{P}}) \rangle_{\mathbb{C}^M} \mathrm{e}^{\frac{2\pi\mathrm{j}pp'}{P}}$$

$$= \{\mathcal{F}^{-1} \langle \mathbf{x}(\mathrm{e}^{\frac{2\pi\mathrm{j}\bullet}{P}}), \boldsymbol{\varphi}(\mathrm{e}^{\frac{2\pi\mathrm{j}\bullet}{P}}) \rangle_{\mathbb{C}^M}\}[p], \quad (8)$$

where "$\bullet$" denotes the variable argument of a given function, and $\mathcal{F}$ is the nonunitary DFT:

$$(\mathcal{F} y)[p] = \sum_{p'=0}^{P-1} y[p'] \mathrm{e}^{\frac{-2\pi\mathrm{j}pp'}{P}}.$$

Taking DFTs of (8) yields the fundamental identity:

$$(\mathcal{F} \langle x, \mathcal{T}^{M\bullet} \varphi \rangle_{\ell(\mathbb{Z}_{MP})})[p] = \langle \mathbf{x}(\mathrm{e}^{\frac{2\pi\mathrm{j}p}{P}}), \boldsymbol{\varphi}(\mathrm{e}^{\frac{2\pi\mathrm{j}p}{P}}) \rangle_{\mathbb{C}^M}. \quad (9)$$

In particular, (9) immediately yields the finite-dimensional version of the well-known polyphase characterization of the orthogonality of a system of translates: letting $x = \varphi_n$ and $\varphi = \varphi_n$, we have that $\{\mathcal{T}^{Mp} \varphi_n\}_{p=0}^{P-1}$ is orthonormal if and only if $\delta_0[p] = \langle \varphi_n, \mathcal{T}^{Mp} \varphi_n \rangle_{\ell(\mathbb{Z}_{MP})}$; by taking DFTs, we see that this occurs precisely when the evaluation of $\boldsymbol{\varphi}(z)$ at any $z = \mathrm{e}^{\frac{2\pi\mathrm{j}p}{P}}$ yields a unit norm vector in $\mathbb{C}^M$. A similar argument shows that the $M$-translates $\varphi_n$ are orthogonal to those of $\varphi_{n'}$ precisely when $\boldsymbol{\varphi}_n(z)$ and $\boldsymbol{\varphi}_{n'}(z)$ are orthogonal in $\mathbb{C}^M$ at any such $z$. Moreover, (8) provides a means for calculating the frame bounds of the synthesis FB $\Phi = \{\mathcal{T}^{Mp} \varphi_n\}_{n=0, p=0}^{N-1, P-1}$:

$$\|\Phi^* x\|_{\ell(\mathbb{Z}_{MP})}^2 = \sum_{n=0}^{N-1} \|\langle x, \mathcal{T}^{M\bullet} \varphi_n \rangle\|_{\ell(\mathbb{Z}_P)}^2$$

$$= \frac{1}{P} \sum_{n=0}^{N-1} \|\mathcal{F} \langle x, \mathcal{T}^{M\bullet} \varphi_n \rangle\|_{\ell(\mathbb{Z}_P)}^2$$

$$= \frac{1}{P} \sum_{p=0}^{P-1} \sum_{n=0}^{N-1} |\langle \mathbf{x}(\mathrm{e}^{\frac{2\pi\mathrm{j}p}{P}}), \boldsymbol{\varphi}_n(\mathrm{e}^{\frac{2\pi\mathrm{j}p}{P}}) \rangle_{\mathbb{C}^M}|^2.$$

In particular, letting $B_p$ be the optimal upper frame bound for $\{\boldsymbol{\varphi}_n(\mathrm{e}^{\frac{2\pi\mathrm{j}p}{P}})\}_{n=0}^{N-1}$ in $\mathbb{C}^M$, we have:

$$\|\Phi^* x\|_{\ell(\mathbb{Z}_{MP})}^2 \leq \frac{1}{P} \sum_{p=0}^{P-1} B_p \|\mathbf{x}(\mathrm{e}^{\frac{2\pi\mathrm{j}p}{P}})\|_{\mathbb{C}^M}^2$$

$$\leq (\max B_p) \|\mathbf{x}(z)\|_{\mathbb{P}_{M,P}}^2$$

$$= (\max B_p) \|x\|_{\ell(\mathbb{Z}_{MP})}^2, \quad (10)$$

with equality holding throughout (10) precisely when $\mathbf{x}(\mathrm{e}^{\frac{2\pi\mathrm{j}p}{P}})$ vanishes for all $p$ such that $B_p$ is not maximal and, at the same time, $\mathbf{x}(\mathrm{e}^{\frac{2\pi\mathrm{j}p}{P}})$ achieves the upper frame bound of $\{\boldsymbol{\varphi}_n(\mathrm{e}^{\frac{2\pi\mathrm{j}p}{P}})\}_{n=0}^{N-1}$ for all maximal $B_p$'s. As such, $\max B_p$ is the optimal upper frame bound for $\{\mathcal{T}^{Mp} \varphi_n\}_{n=0, p=0}^{N-1, P-1}$; a similar derivation provides its optimal lower frame bounds. We summarize the preceding discussion as follows:

**Theorem 1.** *Given filters $\{\varphi_n\}_{n=0}^{N-1}$ in $\ell(\mathbb{Z}_{MP})$:*

(i) *For a fixed $n$, $\{\mathcal{T}^{Mp} \varphi_n\}_{p=0}^{P-1}$ is orthonormal if and only if $\boldsymbol{\varphi}_n(z)$ has unit norm in $\mathbb{C}^M$ at any $z = \mathrm{e}^{\frac{2\pi\mathrm{j}p}{P}}$.*

(ii) *The optimal frame bounds for $\Phi = \{\mathcal{T}^{Mp} \varphi_n\}_{n=0, p=0}^{N-1, P-1}$ in $\ell(\mathbb{Z}_{MP})$ are:*

$$A = \min A_p, \quad B = \max B_p,$$

*where $A_p$ and $B_p$ are the optimal frame bounds for $\{\boldsymbol{\varphi}_n(z)\}_{n=0}^{N-1}$ in $\mathbb{C}^M$ when $z = \mathrm{e}^{\frac{2\pi\mathrm{j}p}{P}}$.*

We emphasize that the novelty of Theorem 1 lies not in the results themselves, as (i) is the finite version of a result used widely throughout the wavelet literature, and (ii) is the finite version of a well-known result on FB frames, namely Corollary 5.1 of [4]. Rather, the novelty lies in the realization that by combining (i) and (ii), one may characterize FBFFs in terms of unit norm frames whose entries are not scalars, but rather, polynomials. In particular, if $\{\varphi_n\}_{n=0}^{N-1}$ satisfies (i) for all $n$, then each channel of the FB $\Phi \Phi^*$, obtained by letting $\tilde{\varphi}_n = \varphi_n$ in Fig. 1, is a projection. That is, in this case, $\Phi \Phi^*$ is the fusion frame operator (6) of the projections:

$$\Pi_n = \sum_{p=0}^{P-1} (\mathcal{T}^{Mp} \varphi_n)(\mathcal{T}^{Mp} \varphi_n)^* = \varphi_n * (\uparrow_M \downarrow_M (\varphi'_n * \bullet)).$$

The optimal frame bounds themselves are given by (ii). The polyphase-based characterizations of Theorem 1 are best visualized in terms of the corresponding $M \times N$ *polyphase matrix* $\Phi(z)$ whose $(m, n)$th entry is the $m$th polyphase component of the $n$th filter:

$$\Phi_{m,n}(z) = \varphi_n^{(m)}(z), \quad (11)$$

and whose $N \times M$ *adjoint* $\Phi^*(z)$ has entries:

$$\Phi^*_{n,m}(z) := [\Phi_{m,n}(z^{-1})]^* = \sum_{p=0}^{P-1} (\varphi_n[m + Mp])^* z^p.$$

Indeed, (i) means that the columns of $\Phi(z)$ have unit norm at each $z = \mathrm{e}^{\frac{2\pi\mathrm{j}p}{P}}$, while the bounds $A_p$ and $B_p$ in (ii) are the least and greatest eigenvalues of $\Phi(z)\Phi^*(z)$ at $z = \mathrm{e}^{\frac{2\pi\mathrm{j}p}{P}}$. In particular, this implies:

**Corollary 2.** *A synthesis FB $\Phi$ generates a tight filter bank fusion frame with each channel of $\Phi\Phi^* = \frac{N}{M}\mathrm{I}$ being a projection if and only if its polyphase matrix $\Phi(z)$ has unit norm columns and orthogonal rows of constant squared-norm $\frac{N}{M}$ at all $z = \mathrm{e}^{\frac{2\pi\mathrm{j}p}{P}}$.*

We note that such polyphase matrices have previously been dubbed *strongly uniform tight* [6]. The remainder of our work is focused almost entirely on the construction of tight FBFFs of the form characterized in Corollary 2, that is, polynomial-entry UNTFs (PUNTFs). We make this restriction due to the fact that such tight, equal-rank fusion frames are known to be optimally robust against certain types of noise and erasures [8], [9]. This restriction also permits us to exploit the existing UNTF literature. The two exceptions to this rule are in Sections IV and V, where we consider tight, nonequal-rank wavelet FBFFs and nontight, equal-rank Gabor FBFFs, respectively. Nevertheless, the following theory represents but a fraction of the FBFFs that can be constructed using Theorem 1. Moreover, Theorem 1 can itself be generalized: by slightly extending an argument of [29], one can in fact unitarily block-diagonalize $\Phi$ in terms of $\Phi(\mathrm{e}^{\frac{2\pi\mathrm{j}p}{P}})$, writing this $MP \times NP$ synthesis operator $\Phi$ in terms of these $P$ matrices of size $M \times N$, as well as DFTs of size $M$, $P$ and $MP$.

### B. Basic Constructions of Filter Bank Fusion Frames

In this section, we present basic methods for constructing PUNTFs, namely polyphase matrices $\Phi(z)$ which satisfy Corollary 2, thereby yielding tight FBFFs. Most simply, one may let $\Phi(z)$ be a constant-entry UNTF $F$; the corresponding analysis FB $\Phi^*$ breaks the signal $x$ in $\ell(\mathbb{Z}_{MP})$ into $P$ nonoverlapping blocks of size $M$, and applies $F^*$ to each.

**Example 3.** If we let $\Phi(z)$ be the popular *Mercedes-Benz* frame of $N = 3$ elements in $M = 2$, namely:

$$\Phi(z) = \frac{1}{2}\left[\begin{array}{ccc} 2 & -1 & -1 \\ 0 & \sqrt{3} & -\sqrt{3} \end{array}\right], \qquad (12)$$

then $\Phi^*$ is the 3-channel, 2-downsampled analysis FB with $\varphi_0 = \delta_0$, $\varphi_{1,2} = \frac{1}{2}(-\delta_0 \pm \sqrt{3}\delta_1)$. Since $\Phi(z)$ is a UNTF — unit-norm columns and orthogonal, constant-norm rows— for every $z$, then by Corollary 2, the FB $\Phi\Phi^*$ consists of three projections of rank $\frac{P}{2}$ which sum to $\frac{3}{2}\mathrm{I}$, namely those projections onto the even translates of $\varphi_0$, $\varphi_1$ and $\varphi_2$.

Such PUNTFs are not applicable to most real-world problems, as they provide little freedom for designing filters with good frequency responses, and also induce blocking artifacts in reconstructed data. Nevertheless, the fact that UNTFs correspond to constant PUNTFs inspires one to see which of the very few known methods for constructing UNTFs extend to the PUNTF setting. Indeed, most of these methods will extend, provided they do not involve division: one may add, subtract and multiply the entries of polyphase matrices in the usual manner, provided they all lie in the same ring $\mathbb{C}[z]/\langle z^P - 1\rangle$, that is, provided the periods of the corresponding filters are chosen consistently; division is trickier, and we avoid it; even if one alternatively considers FBs over $\ell^2(\mathbb{Z})$, dividing FIR polyphase components often leads to unappetizing infinite impulse response filters. Specifically, we have:

**Proposition 4.**

(i) *If $\Phi_0(z)$ and $\Phi_1(z)$ are $M \times N_0$ and $M \times N_1$ PUNTFs, respectively, then so is their $M \times (N_0 + N_1)$ union $[\ \Phi_0(z)\ |\ \Phi_1(z)\ ]$.*

(ii) *If $\Phi_0(z)$ and $\Phi_1(z)$ are $M_0 \times N_0$ and $M_1 \times N_1$ PUNTFs, respectively, then so is their $M_0M_1 \times N_0N_1$ tensor product $\Phi_0(z) \otimes \Phi_1(z)$.*

(iii) *If $\Psi(z)$ is an $M \times M$ unitary polyphase matrix and $\Phi(z)$ is an $M \times N$ PUNTF, then $\Psi(z)\Phi(z)$ is also an $M \times N$ PUNTF.*

The proofs of these elementary results are omitted. Note that (i) corresponds to simply stacking the channels of two FBFFs with identical downsampling rates into a single, larger FB. A similar intuitive understanding of (ii) remains elusive: unlike wavelet constructions, there does not seem to be a clean expression for the frequency responses of the filters defined by $\Phi_0(z) \otimes \Phi_1(z)$ in terms of those defined by $\Phi_0(z)$ and $\Phi_1(z)$. Meanwhile, (iii) was previously used to construct PUNTFs in [6] in the special case where $\Phi(z)$ was constant. Indeed, this approach allows one to make use of extensive wavelet literature of paraunitary matrices, and moreover, of the fact that all such matrices can be factored into multiplicands of the form $(\mathrm{I} - uu^*) + zuu^*$ where $u$ is a unit vector in $\mathbb{C}^M$ [27]:

**Example 5.** Let $F$ be the Mercedes-Benz frame (12), and let $\Psi(z)$ be the unitary polyphase matrix of the 2-channel, 4-tap, 2-downsampled Daubechies FB [30]:

$$\Psi(z) = \left[\begin{array}{cc} a + bz^{-1} & d + cz^{-1} \\ c + dz^{-1} & -b - az^{-1} \end{array}\right], \qquad (13)$$

where $a, d = 2^{-\frac{5}{2}}(1 \pm \sqrt{3})$, $b, c = 2^{-\frac{5}{2}}(3 \mp \sqrt{3})$. As such, the set of even translates of the corresponding low- and high-pass filters $\psi_0 = a\delta_0 + c\delta_1 + b\delta_2 + d\delta_3$ and $\psi_1 = d\delta_0 - b\delta_1 + c\delta_2 - a\delta_3$, form an orthonormal basis for $\ell(\mathbb{Z}_{2P})$. Meanwhile, by (iii) of Proposition 4, the matrix $\Phi(z)$ built as:

$$\Psi(z)F = \left[\begin{array}{ccc} a + bz^{-1} & -2^{-\frac{3}{2}}(1 - \sqrt{3}z^{-1}) & d - cz^{-1} \\ c + dz^{-1} & -2^{-\frac{3}{2}}(\sqrt{3} + z^{-1}) & -b + az^{-1} \end{array}\right]$$

is a PUNTF, implying, by Corollary 2, that the corresponding 3-channel, 4-tap, 2-downsampled FB is a tight FBFF with each channel being a projection. That is, the even translates of the corresponding filters $\{\varphi_0, \varphi_1, \varphi_2\}$ form a $\frac{3}{2}$-tight frame for $\ell(\mathbb{Z}_{2P})$, with the even translates of any particular $\varphi_n$ being orthonormal. The squared magnitude of the frequency responses $|\hat{\varphi}_n(\omega)|^2 = |\sum_k \varphi_n[k]\mathrm{e}^{-\mathrm{j}k\omega}|^2$ of these three real-valued filters are depicted in Fig. 3. As the first columns of $\Phi(z) = \Psi(z)F$ and $\Psi(z)$ are equal, $\varphi_0$ is the traditional 4-tap low-pass Daubechies filter $\psi_0$. Meanwhile, $\varphi_1, \varphi_2 = \frac{1}{2}(-\psi_0 \pm \sqrt{3}\psi_1)$, being combinations of low- and high-pass filters, exhibit poor frequency selectivity. As such, this FB $\Phi$, though a mathematically valid example of a FBFF, is not well-suited for many signal processing applications.

Similar problems befall many 3-channel, 2-downsampled tight FBFFs: the Mercedes-Benz frame is, up to rotation and unit-scalar multiplications, the only $2 \times 3$ UNTF [15], and so any $2 \times 3$ PUNTF $\Phi(z)$ can be essentially factored as $\Phi(z) = \Psi(z)F$, where $\Psi(z)$ is unitary; the three filters of $\Phi$ are therefore nontrivial linear combinations of the two filters of $\Psi$, and as such, are likely to have poor frequency responses and no good coverage of the frequency domain. One solution to this problem is to increase the number of filters, and, perhaps, the downsampling rate. In particular, in the next section, we construct 4-channel, 2-downsampled FBFFs: as any $2 \times 4$ UNTF consists of two orthonormal bases, we construct a $2 \times 4$ PUNTF by applying Proposition 4(i) where $\Phi_0(z)$ and $\Phi_1(z)$ are modulated copies of (13), see Example 7 below. Moreover, as such FBs correspond to combining two one-level wavelet transforms, we are naturally led to investigate how, in a manner similar to that of multi-level wavelet transforms, FBFFs can be linked together to produce larger ones.

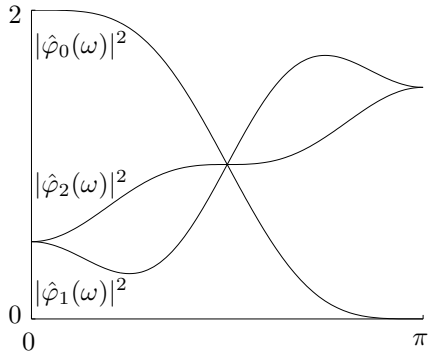

Fig. 3. The frequency responses of three filters of the 3-channel, 4-tap, 2-downsampled tight filter bank fusion frame constructed in Example 5. This filter bank meets some, but not all, of our criteria for nice filter banks: being a tight filter bank fusion frame, it is robust with respect to noise and erasures; moreover, it operates quickly, as each filter has only 4-taps; but while $\varphi_0$ is purely low-pass, $\varphi_1$ and $\varphi_2$ have poor frequency selectivity.

## IV. DISCRETE WAVELET FUSION FRAMES

In this section, we show how by chaining together the tight FBFFs of the previous section, we can produce fusion frame versions of unitary discrete wavelet transforms (DWTs). Here, it is necessary to generalize the notion of a fusion frame to that of a *weighted fusion frame* [7] in which the summands of the fusion frame operator (6) are accompanied by nonnegative weights. Moreover, as we are chiefly concerned with tight frames, and may always absorb the tightness constant into the weights, we assume the tightness constant is 1. To be precise, letting $\mathbb{H}$ be a Hilbert space, we say that a sequence of operators $\{F_k : \ell(\mathbb{I}_k) \to \mathbb{H}\}_{k \in \mathbb{K}}$ *induces* a *Parseval fusion frame* with nonnegative *weights* $\{c_k\}_{k \in \mathbb{K}}$ if $F_k^* F_k = \mathrm{I}$ for all $k \in \mathbb{K}$ (or equivalently, that each $\Pi_k = F_k F_k^*$ is an orthogonal projection) and:

$$\sum_{k \in \mathbb{K}} c_k \Pi_k = \sum_{k \in \mathbb{K}} c_k F_k F_k^* = \mathrm{I}.$$

The following result shows how such fusion frames can be combined to create new ones:

**Theorem 6.** *Let* $\{F_k : \ell(\mathbb{I}_k) \to \mathbb{H}\}_{k \in \mathbb{K}}$ *induce a Parseval fusion frame for* $\mathbb{H}$ *with weights* $\{c_k\}_{k \in \mathbb{K}}$*, and for each* $k \in \mathbb{K}$*, let* $\{G_{k,l}\}_{l \in \mathbb{L}_k}$ *induce a Parseval fusion frame for* $\ell(\mathbb{I}_k)$ *with weights* $\{d_{k,l}\}_{l \in \mathbb{L}_k}$*. Then,* $\{F_k G_{k,l}\}_{k \in \mathbb{K}, l \in \mathbb{L}_k}$ *induces a Parseval fusion frame for* $\mathbb{H}$ *with weights* $\{c_k d_{k,l}\}_{k \in \mathbb{K}, l \in \mathbb{L}_k}$*.*

The proof of Theorem 6 is simple, noting:

$$(F_k G_{k,l})^*(F_k G_{k,l}) = G_{k,l}^* F_k^* F_k G_{k,l} = G_{k,l}^* \mathrm{I} G_{k,l} = \mathrm{I},$$

for any $k \in \mathbb{K}$ and $l \in \mathbb{L}_k$, while:

$$\sum_{k \in \mathbb{K}} \sum_{l \in \mathbb{L}_k} c_k d_{k,l} (F_k G_{k,l})(F_k G_{k,l})^*$$
$$= \sum_{k \in \mathbb{K}} c_k F_k \Big( \sum_{l \in \mathbb{L}_k} d_{k,l} G_{k,l} G_{k,l}^* \Big) F_k^*$$
$$= \sum_{k \in \mathbb{K}} c_k F_k \mathrm{I} F_k^*$$
$$= \mathrm{I}.$$

The significance of Theorem 6 is that it permits us to build intricate FBFFs by composing the operators of simpler ones, namely discrete wavelet FBFFs. Though the ideas presented here can produce DWTs of any number of levels, we, for the sake of clarity, consider only two-level DWTs. To be precise, let $\{\varphi_n\}_{n=0}^{N-1}$ be the filters in an $M$-downsampled tight FBFF over $\ell(\mathbb{Z}_{M^2 P})$ of the type characterized in Corollary 2. In the parlance of Theorem 6, we have $\mathbb{H} = \ell(\mathbb{Z}_{M^2 P})$, with each $\mathbb{I}_k = \mathbb{Z}_{MP}$ and each $F_k = \Phi_n = \{\mathcal{T}^{Mq} \varphi_n\}_{q=0}^{MP-1}$ corresponding to a single channel of the synthesis FB; viewed as operators, we have:

$$\Phi_n : \ell(\mathbb{Z}_{MP}) \to \ell(\mathbb{Z}_{M^2 P}), \quad \Phi_n y = \varphi_n * (\uparrow_M y).$$

As the $M$-translates of $\varphi_n$ are orthonormal for any single $n$, we have $\Phi_n^* \Phi_n = \mathrm{I}$. Moreover, as $\frac{N}{M} \mathrm{I} = \Phi \Phi^* = \sum \Phi_n \Phi_n^*$, we have that $\{\Phi_n\}_{n=0}^{N-1}$ induces a Parseval fusion frame with constant weights $c_n = \frac{M}{N}$. Next, we choose the $G_{k,l}$'s; the manner in which we do so determines the tree structure of our fusion frame DWT. In particular, to build a fusion frame version of wavelet packets — in the second level, each channel is passed through a copy of the first-level FB — we choose all the $G_{k,l}$'s to be copies of $\Phi_n$'s with $d_{k,l} = \frac{M}{N}$. If, on the other hand, a traditional DWT tree is desired — the 0th channel is passed through a copy of the first-level FB, while the other $N - 1$ channels are untouched — then the $G_{0,l}$'s are copies of $\Phi_n$'s with $d_{0,l} = \frac{M}{N}$, while the remaining $G_{k,l}$'s essentially do not exist; formally, for $k \geq 1$, we let $\mathbb{L}_k = \{0\}$, $G_{k,0} = \mathrm{I}$ and $d_{k,0} = 1$. A technical point: one may not choose the second level filters to be exactly identical to the first, as they lie in different spaces, namely $\ell(\mathbb{Z}_{MP})$ and $\ell(\mathbb{Z}_{M^2 P})$, respectively; rather, they should be chosen as their *periodizations*; for $\varphi_n \in \ell(\mathbb{Z}_{M^2 P})$, consider:

$$\mathring{\varphi}_n \in \ell(\mathbb{Z}_{MP}), \quad \mathring{\varphi}_n[k] := \sum_{m=0}^{M-1} \varphi_n[k + MPm],$$

whose polyphase components satisfy $\mathring{\varphi}_n^{(m)}(z) = \varphi_n^{(m)}(z)$ for all $z = \mathrm{e}^{\frac{2\pi \mathrm{j} p}{P}}$; as such, if $\Phi(z)$ satisfies Corollary 2, then $\mathring{\Phi}(z)$ does as well; filters with low numbers of taps are essentially equal to their periodizations, and we ignore this subtle distinction from this point forward. We conclude this section with an example of discrete wavelet FBFFs.

**Example 7.** As the $2 \times 3$ PUNTFs discussed in Example 5 yielded filters with poor frequency selectively, we focus here on $2 \times 4$ PUNTFs, that is, 4-channel, 2-downsampled tight FBFFs $\Phi$ in which each channel of $\Phi\Phi^*$ is a projection. As any UNTF of four elements in two-dimensional space consists of two orthonormal bases [15], we consider the PUNTFs given by Proposition 4(i), namely $\Phi(z) = [\ \Phi_0(z)\ |\ \Phi_1(z)\ ]$, where $\Phi_0(z)$ and $\Phi_1(z)$ are unitary $2 \times 2$ polyphase matrices. That is, we consider the FBs obtained by stacking two orthogonal DWTs. For example, we can let $\Phi_0(z)$ be the polyphase matrix $\Psi(z)$ of the 4-tap Daubechies DWT, as given in Example 5. Though we may also take $\Phi_1(z)$ to be $\Psi(z)$, a more interesting example is obtained by modulating the $\psi$'s, shifting their frequencies by $\frac{\pi}{2}$. That is, letting $\Phi_0(z) = \Psi(z)$ and $\Phi_1(z) = [\begin{smallmatrix}1 & 0\\0 & j\end{smallmatrix}]\Psi(-z)$ yields the $2 \times 4$ PUNTF:

$$\begin{bmatrix} a+bz^{-1} & d+cz^{-1} & a-bz^{-1} & d-cz^{-1} \\ c+dz^{-1} & -b-az^{-1} & \mathrm{j}c-\mathrm{j}dz^{-1} & -\mathrm{j}b+\mathrm{j}az^{-1} \end{bmatrix}.$$

The frequency responses of the corresponding filters $\{\varphi_n\}_{n=0}^3$ are depicted in Fig. 4(a), with a diagram of the corresponding synthesis FB $\Phi$ given in Fig. 4(d). The corresponding tight FBFF has redundancy 2, decomposing $4Q$-dimensional signals in terms of 4 signals of dimension $2Q$. Note that such a FBFF corresponds to a weighted Parseval frame, of uniform weight $c_n = \frac{1}{2}$. We further note that in this particular example, we additionally have that the zeroth and first channels are mutually orthogonal, as are the second and third, implying the corresponding FBFF actually breaks such signals into two copies of themselves; we do not pursue this fact further here, as it does not necessarily generalize to higher-redundancy FBFFs.

The remainder of Fig. 4 depicts two distinct ways in which, using Theorem 6, this FBFF may be iterated so as to produce more intricate FBFFs. To be precise, the second and third columns of Fig. 4 depict FBFF generalizations of traditional DWTs and wavelet packets, respectively. Here, since $\varphi_n * (\uparrow_M (\varphi_{n'} * y)) = (\varphi_n * (\uparrow_M \varphi_{n'})) * (\uparrow_M y)$, we may consider the frequency responses of the *equivalent filter* of any channel: $\varphi_{n,n'} := \varphi_n * (\uparrow_M \varphi_{n'})$; one of the true advantages of employing such iterated FBs is that the frequency-responses of their equivalent filters are predictable: $|\varphi_{n,n'}(t)|^2 = |\varphi_n(t)|^2 |\varphi_{n'}(Mt)|^2$. In particular, if only the low-pass channel is again passed through $\Phi$, à la Fig. 4(e), the corresponding 2-level transform is equivalent to a 7-channel FB whose frequency responses are depicted in Fig. 4(b). Here, the four channels corresponding to $\{\varphi_{0,0}, \varphi_{0,1}, \varphi_{0,2}, \varphi_{0,3}\}$ have a downsampling rate of 4, while the three channels corresponding to $\{\varphi_1, \varphi_2, \varphi_3\}$ have a downsampling rate of 2. By Theorem 6, this FB corresponds to a weighted Parseval fusion frame which breaks $4Q$-dimensional signals into 7 components, namely its projections onto four $Q$-dimensional subspaces, each weighted by $\frac{1}{4}$, and three $2Q$-dimensional subspaces, each weighted by $\frac{1}{2}$. If, on the other hand, each of the four original channels is again passed through $\Phi$, the resulting wavelet packet, depicted in Fig. 4(c),(f), is a full tree and breaks $4Q$-dimensional signals into sixteen $Q$-dimensional projections, each weighted by $\frac{1}{4}$.

The preceding example highlights the great control in frequency design offered by the use of modulation. In the following section, we focus exclusively on FBs whose filters all arise as the modulates of a single filter.

## V. DISCRETE GABOR FUSION FRAMES

In this section, we characterize and construct *Gabor* FBFFs. That is, we consider the fusion frame properties of collections of translates and *modulates* of a single filter; for $y \in \ell(\mathbb{Z}_P)$, let $(\mathcal{M}^p y)[q] := \mathrm{e}^{\frac{2\pi\mathrm{j}pq}{P}} y[q]$ by the modulation of $y$ by a factor of $p$. Here, we assume that the number of channels $N$ is divisible by the downsampling rate $M$; this integer-redundancy case is significantly less complicated than the noninteger case [31]. In particular, letting $N = MR$, $P = QR$ and $\varphi \in \ell(\mathbb{Z}_{MQR})$, we consider the filters $\{\varphi_n\}_{n=0}^{MR-1}$, where $\varphi_n[k] = (\mathcal{M}^{Qn}\varphi)[k] = \mathrm{e}^{\frac{2\pi\mathrm{j}kn}{MR}}\varphi[k]$. That is, we consider the fusion frame properties of the system $\{\mathcal{T}^{Mp}\mathcal{M}^{Qn}\varphi\}_{p=0,\ n=0}^{QR-1\ MR-1}$ of redundancy $R$. This corresponds exactly to an $MR$-channel, $M$-downsampled FB whose filters are regular modulates of a single template $\varphi$.

Letting $\boldsymbol{\varphi}(z)$ be the polyphase vector (7) of $\varphi$, one may easily show that the $m$th polyphase component of $\varphi_n$ is $\mathrm{e}^{\frac{2\pi\mathrm{j}mn}{MR}}\varphi^{(m)}(\mathrm{e}^{\frac{-2\pi\mathrm{j}n}{R}}z)$. This fact immediately implies that the rows of the corresponding polyphase matrix $\Phi(z)$ are orthogonal: letting $n = r + Rk$,

$$(\Phi(z)\Phi^*(z))_{m,m'} = \sum_{k=0}^{M-1} \mathrm{e}^{\frac{2\pi\mathrm{j}(m-m')k}{M}}$$
$$\times \sum_{r=0}^{R-1} \mathrm{e}^{\frac{2\pi\mathrm{j}(m-m')r}{MR}} \varphi^{(m)}(\mathrm{e}^{\frac{-2\pi\mathrm{j}r}{R}}z)(\varphi^{(m')}(\mathrm{e}^{\frac{-2\pi\mathrm{j}r}{R}}z))^*;$$

evaluating the geometric sum, we see that $\Phi(z)\Phi^*(z)$ is diagonal. Its eigenvalues are thus obtained by letting $m = m'$; Theorem 1 then recovers a finite-dimensional version of a previously known result [28] giving frame bounds on integer-redundancy Gabor systems in terms of their Zak transforms:

**Theorem 8.** *Given $\varphi \in \ell(\mathbb{Z}_{MQR})$, the optimal frame bounds of $\{\mathcal{T}^{Mp}\mathcal{M}^{Qn}\varphi\}_{p=0,\ n=0}^{QR-1\ MR-1}$ are the extreme values of:*

$$M \sum_{r=0}^{R-1} |\varphi^{(m)}(\mathrm{e}^{\frac{-2\pi\mathrm{j}r}{R}}z)|^2$$

*over all $m = 0, \ldots, M-1$ and all $z = \mathrm{e}^{\frac{2\pi\mathrm{j}p}{QR}}$ with $p \in \mathbb{Z}$.*

In light of Theorem 8, we define the $M \times R$ *Zak matrix* $\Phi^{\mathrm{Zak}}(z)$ of $\varphi$ to be the first $R$ columns of its $M \times MR$ polyphase matrix $\Phi(z)$, sans the external modulation factors:

$$\Phi^{\mathrm{Zak}}_{m,r}(z) = \varphi^{(m)}(\mathrm{e}^{\frac{-2\pi\mathrm{j}r}{R}}z).$$

In particular, Theorem 8 gives that the synthesis FB $\Phi$ satisfies $\Phi\Phi^* = A\mathrm{I}$ if and only if the rows of $\Phi^{\mathrm{Zak}}(z)$ are of constant norm. Moreover, as shown below, Theorem 1 implies that each channel of $\Phi\Phi^*$ is a projection if and only if the columns of $\Phi^{\mathrm{Zak}}(z)$ are of constant norm. Interestingly, when $\Phi$ is tight but each channel is not a projection — the rows of $\Phi^{\mathrm{Zak}}(z)$ are of constant norm, but the columns are not — each channel is nevertheless a sum of $R$ projections. That is, if $\varphi$ generates a tight $MR$-channel, $M$-downsampled Gabor FB, then it is necessarily a FBFF, consisting of $MR^2$ rank-$Q$ projections:

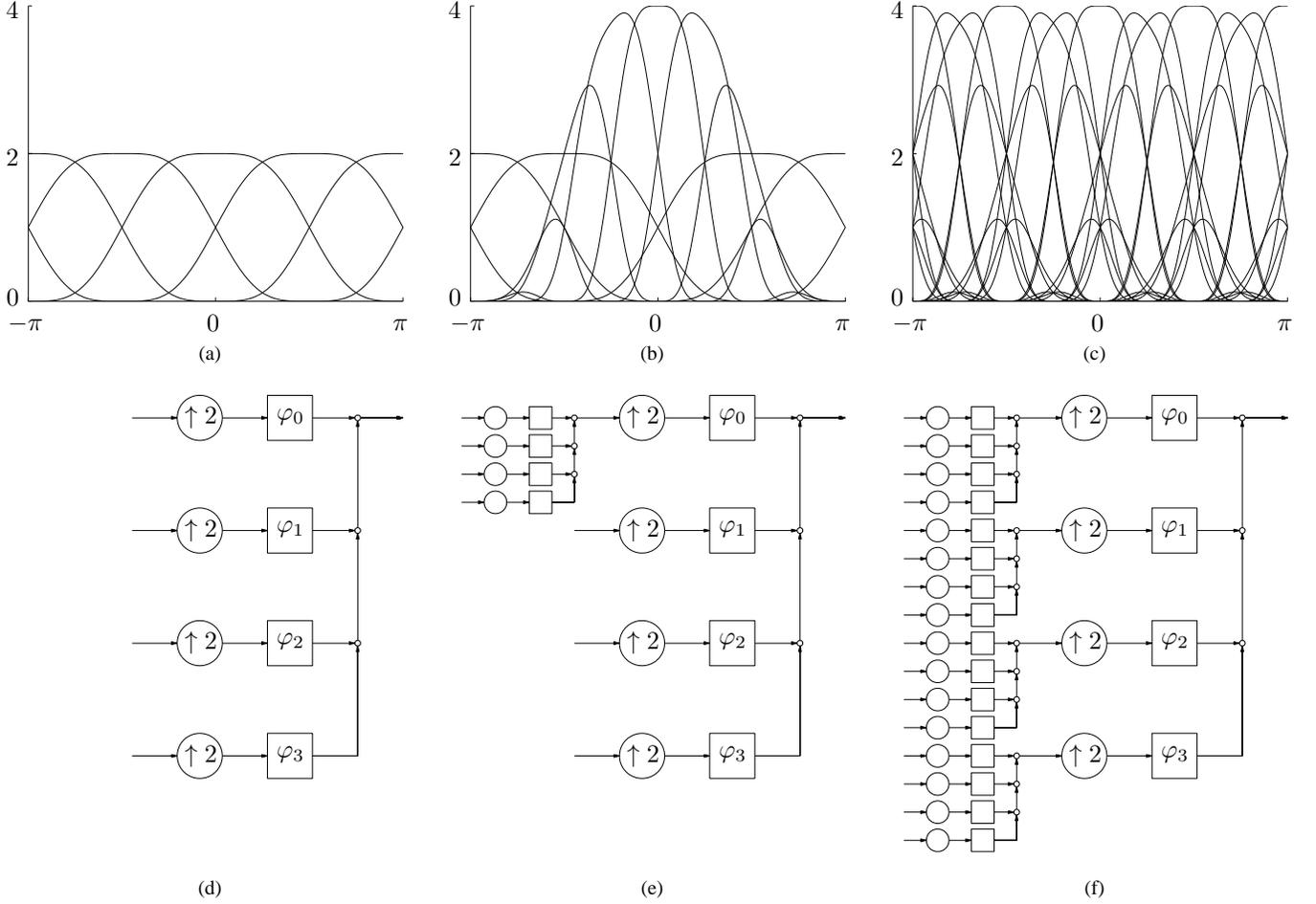

Fig. 4. The tight filter bank fusion frames constructed in Example 7 using Theorem 6. The first column, from top to bottom, presents the frequency-responses and synthesis filter bank diagram, respectively, of the 4-channel, 2-downsampled tight filter bank fusion frame obtained by combining a traditional discrete wavelet transform with one of its modulates. The corresponding fusion frame has redundancy two, breaking signals into four half-dimensional components. In the second column, in a manner analogous to traditional discrete wavelet transforms, the low-pass channel is passed through the filter bank a second time, producing a new tight filter bank fusion frame, which decomposes signals in terms of seven projections: four of quarter dimension, and three of half dimension. In the third column, every channel is again passed through the filter bank; similar to wavelet packets, the corresponding tight filter bank fusion frame breaks signals into sixteen quarter-dimensional components.

**Theorem 9.** *The Gabor sequence* $\{\mathcal{T}^{Mp}\mathcal{M}^{Qn}\varphi\}_{p=0,\ n=0}^{QR-1\ MR-1}$ *of a unit norm* $\varphi \in \ell(\mathbb{Z}_{MQR})$:

(i) *has each subsequence* $\{\mathcal{T}^{Mp}\mathcal{M}^{Qn}\varphi\}_{p=0}^{QR-1}$ *being orthonormal if and only if:*

$$\sum_{m=0}^{M-1} |\varphi^{(m)}(z)|^2 = 1,$$

*for all* $z = \mathrm{e}^{\frac{2\pi\mathrm{j}p}{QR}}$;

(ii) *is a tight frame if and only if:*

$$\sum_{r=0}^{R-1} |\varphi^{(m)}(\mathrm{e}^{\frac{-2\pi\mathrm{j}r}{R}}z)|^2 = \frac{R}{M} \quad (14)$$

*for all* $m = 0,\ldots,M-1$ *and all* $z = \mathrm{e}^{\frac{2\pi\mathrm{j}p}{QR}}$, *or equivalently when, for every fixed* $m$, *the* $R$-*translates of* $\sqrt{M}\varphi[m + M\bullet]$ *are orthonormal.*

*Moreover, in this case,* $\{\mathcal{T}^{M(r+Rq)}\mathcal{M}^{Qn}\varphi\}_{q=0}^{Q-1}$ *is necessarily an orthonormal set for any fixed* $n$ *and* $r$.

The proof of this result is given in the appendix. A characterization of tightness closely related to (ii) is given in [31]. Note that in the special case where $M = R$, then for any $\varphi^{(0)}$ that satisfies (14), namely:

$$\sum_{m=0}^{M-1} |\varphi^{(0)}(\mathrm{e}^{\frac{-2\pi\mathrm{j}m}{M}}z)|^2 = 1, \quad \forall z = \mathrm{e}^{\frac{2\pi\mathrm{j}p}{MQ}}, \quad (15)$$

letting $\varphi^{(m)}(z) = \varphi^{(0)}(\mathrm{e}^{\frac{-2\pi\mathrm{j}m}{M}}z)$ for all $m = 1,\ldots,M-1$, yields a $\varphi$ which satisfies all of the conditions of Theorem 9. We conclude this section with an example, using Theorem 8 to construct a 4-channel, 2-downsampled tight Gabor FBFF:

**Example 10.** Let $M = R = 2$ and let $Q$ be arbitrary. Though the full $2 \times 4$ polyphase matrix is:

$$\Phi(z) = \begin{bmatrix} \varphi^{(0)}(z) & \varphi^{(0)}(-z) & \varphi^{(0)}(z) & \varphi^{(0)}(-z) \\ \varphi^{(1)}(z) & \mathrm{j}\varphi^{(1)}(-z) & -\varphi^{(1)}(z) & -\mathrm{j}\varphi^{(1)}(-z) \end{bmatrix},$$

we see that the rows of this matrix are automatically orthogonal, as guaranteed in general by the proof of Theorem 8. As

such, we need only consider the smaller $2 \times 2$ Zak matrix:
$$\Phi^{\text{Zak}}(z) = \begin{bmatrix} \varphi^{(0)}(z) & \varphi^{(0)}(-z) \\ \varphi^{(1)}(z) & \varphi^{(1)}(-z) \end{bmatrix}.$$

In particular, Theorem 9 gives that the discrete Gabor system $\{\mathcal{T}^{2p}\mathcal{M}^{Qn}\varphi\}_{p=0,\ n=0}^{2Q-1,\ 3}$ arising from some unit norm $\varphi$ in $\ell(\mathbb{Z}_{4Q})$ is tight when:

$$|\varphi^{(0)}(z)|^2 + |\varphi^{(0)}(-z)|^2 = 1, \quad (16)$$
$$|\varphi^{(1)}(z)|^2 + |\varphi^{(1)}(-z)|^2 = 1. \quad (17)$$

Equivalently, this system is tight if both the even and odd parts of $\varphi$ have norm $2^{-\frac{1}{2}}$ and are each orthogonal to their own even translates. In wavelet terminology, we need both the even and odd parts of $\varphi$ to be admissible discrete scaling functions. For this to hold, the 4-translates of $\varphi$ are necessarily orthogonal, as are the 4-translates of $\mathcal{T}^2\varphi$ — each channel is a sum of two projections of rank $Q$ — yielding a FBFF of eight rank-$Q$ projections. Note that if we further have:

$$|\varphi^{(0)}(z)|^2 + |\varphi^{(1)}(z)|^2 = 1, \quad (18)$$

then Theorem 9 gives that the 2-translates of $\varphi$ are orthonormal — each channel is a single projection of rank $2Q$ — yielding a FBFF of four rank-$2Q$ projections. As noted above, one way to ensure (18) is to choose any $\varphi^{(0)}(z)$ that satisfies (16) and let $\varphi^{(1)}(z) = \varphi^{(0)}(-z)$. Seeking the greatest flexibility with regards to the frequency response of $\varphi$, we, from this point forward, do not make the additional restriction (18), settling for the weaker fusion properties guaranteed by (16) and (17). Indeed, inspired by the max-flat method for constructing discrete Daubechies wavelets [30], we seek those $T$-degree polynomials $\varphi^{(0)}(z)$ and $\varphi^{(1)}(z)$ which satisfy (16) and (17) whose corresponding filter $\varphi$ has the flattest possible frequency response at the origin. Specifically, we want the first $T$ derivatives of:

$$\sum_{k=0}^{2T-1} \varphi[k]z^k = \sum_{p=0}^{T-1} \varphi[2p]z^{2p} + \sum_{p=0}^{T-1} \varphi[2p+1]z^{2p+1}$$

to vanish at $z = 1$. This equates to a linear relationship between the coefficients of $\varphi^{(0)}(z)$ and those of $\varphi^{(1)}(z)$:

$$\sum_{p=\lceil\frac{k-1}{2}\rceil}^{T-1} \frac{(2p+1)!}{(2p+1-k)!}\varphi[2p+1] = -\sum_{p=\lceil\frac{k}{2}\rceil}^{T-1} \frac{(2p)!}{(2p-k)!}\varphi[2p],$$

for all $k = 0, \ldots, T-1$. As such, we can use standard nonlinear solvers, such as MATLAB's `fsolve`, to attempt to find particular values of the independent variables $\{\varphi[2p]\}_{p=0}^{T-1}$ whose corresponding $\varphi_0(z)$ and $\varphi_1(z)$ satisfy the Gabor tightness conditions (16) and (17). The numerical evidence indicates that these equations have a nonempty, discrete set of solutions when $T$ is even. A 20-tap max-flat filter $\varphi$ satisfying (16) and (17) with $T = 10$ is depicted in Fig. 5.

## VI. CONCLUSIONS

We have characterized and constructed perfect reconstruction FBFFs with desirable properties making them stable and robust to noise. We focused on the case where each channel of the FB is an orthogonal projection. We used polyphase representations to characterize FBFFs and proved that tight FBFFs correspond to polyphase matrices with unit-norm columns and orthogonal rows of constant norm. Additionally, we presented various ways of constructing new FBFFs. In particular, we constructed tight FBFFs with wavelet-like structures. Finally we studied Gabor fusion frames and designed Gabor FBFF filters that have good localization in both time and frequency.

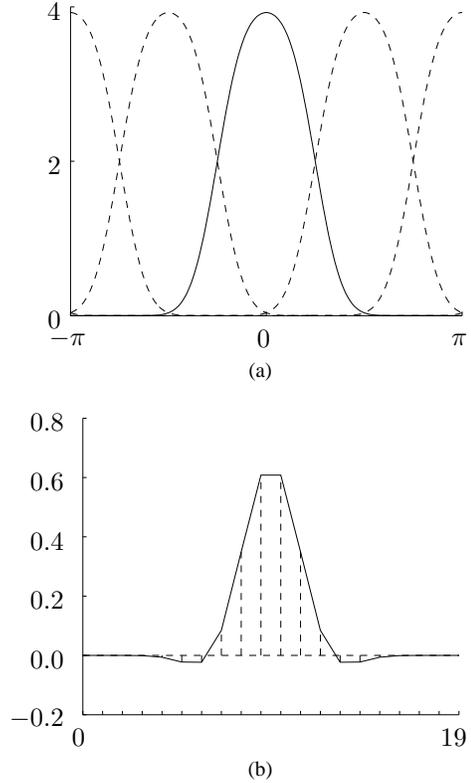

Fig. 5. A 20-tap max flat filter $\varphi$ whose four modulates and even translates form a 4-channel, 2-downsampled tight filter bank fusion frame. The frequency response of $\varphi$ and its three modulates are depicted in (a), with the filter itself depicted in (b). The 2-translates of $\varphi$ are not orthogonal, and so, unlike the preceding examples, each channel of the filter bank does not correspond to a projection. Nevertheless, in accordance with Theorem 9(b), the 4-translates of $\varphi$ are necessarily orthonormal. Indeed, each channel of the filter bank corresponds to a sum of two projections, for a total of eight quarter-dimensional projections. The fact that this particular filter is even appears to be a happy coincidence; most lower-tap analogs do not possess such symmetry.

## APPENDIX A
## PROOF OF THE THEOREM 9

For (i), note that by Theorem 1, the assumption is equivalent to having $\{\mathcal{T}^{Mp}\varphi\}_{p=0}^{QR-1}$ being orthonormal. When coupled with the following relation:

$$\langle \mathcal{M}^{Qn}\varphi, \mathcal{T}^{Mp}\mathcal{M}^{Qn}\varphi\rangle = e^{\frac{2\pi\mathrm{j}np}{R}}\langle \varphi, \mathcal{T}^{Mp}\varphi\rangle, \quad (19)$$

this fact is immediately equivalent to having each $\{\mathcal{T}^{Mp}\mathcal{M}^{Qn}\varphi\}_{p=0}^{QR-1}$ being orthonormal.

For (ii), the first characterization of tightness (14) is an immediate consequence of Theorem 8 and the fact that the tight frame constant $A$ of this UNTF is necessarily its redundancy $R$. For the second characterization, plugging (7) into (14) with $p = r' + Rq$ gives that tightness is equivalent to having:

$$\sum_{r=0}^{R-1}\left|\sum_{r'=0}^{R-1}\left(\sum_{q=0}^{Q-1}\varphi[m+Mr'+MRq]z^{-r'-Rq}\right)e^{\frac{2\pi j r r'}{R}}\right|^2 = \frac{R}{M}$$

for all $m = 0, \ldots, M-1$ and all $z = e^{\frac{2\pi j p}{QR}}$. As the $R \times R$ DFT satisfies $\mathcal{F}^*\mathcal{F} = R\mathrm{I}$, this, in turn, is equivalent to:

$$\sum_{r=0}^{R-1}\left|\sum_{q=0}^{Q-1}\sqrt{M}\varphi[m+M(r+Rq)](z^R)^{-q}\right|^2 = 1,$$

that is, to having the $R$-downsampled polyphase vectors of each $\sqrt{M}\varphi[m+M\bullet]$ having unit norm at all $w = z^R = e^{\frac{2\pi j p}{Q}}$. By Theorem 1, tightness is therefore equivalent to having the $R$-translates of each $\sqrt{M}\varphi[m+M\bullet]$ being orthonormal. This fact then immediately implies that the $MR$-translates of $\varphi$ are orthonormal, which in turn, by (19), implies that each subsequence $\{\mathcal{T}^{M(r+Rq)}\mathcal{M}^{Qn}\varphi\}_{q=0}^{Q-1}$ is orthonormal.

## ACKNOWLEDGMENTS

The authors would like to thank Martin Vetterli for helpful discussions.